\documentclass{article}

\usepackage[english]{babel}
\usepackage{braket}
\usepackage{subfigure}
\usepackage[numbers,sort&compress]{natbib}
\usepackage[letterpaper,top=2cm,bottom=2cm,left=3cm,right=3cm,marginparwidth=1.75cm]{geometry}

\usepackage{amssymb}
\usepackage{amsmath}
\usepackage{graphicx}
\usepackage[colorlinks=true, allcolors=blue]{hyperref}

\title{Enhanced Sensitivity in Rydberg Atom Electric Field Sensors through Autler-Townes Effect and Two-Photon Absorption: A Theoretical Analysis Using Many-Mode Floquet Theory}
\author{Tianhao Wu \thanks{Tianhao Wu is with the School of Physics, Xi'an Jiaotong University, Xi'an, Shaanxi 710049, China (e-mail: wutianhao\_quantum@stu.xjtu.edu.cn).}}

\begin{document}
\maketitle

\begin{abstract}
In this paper, we present a comprehensive investigation into the sensitivity of a Rydberg atom electric field sensor, with a specific focus on the minimum detectable field (MDF) as a key metric. The study utilizes one-mode Floquet theory to calculate the Stark shift for selected Rydberg states when exposed to a signal electric field. The results are compared to those obtained using the rotating wave approximation (RWA). To enhance the sensor's sensitivity when the frequency of the signal electric field deviates from resonance frequencies between Rydberg states, we propose incorporating an extra coupling electric field and using many-mode Floquet theory, a generalization of one-mode Floquet theory, to theoretically analyze this kind of Rydberg atom electric field sensor. The Autler-Townes effect resulting from this coupling electric field causes Rydberg states to split into dressed states, effectively increasing sensitivity by modulating the frequencies of resonance peaks. Moreover, the phenomenon of two-photon absorption in the presence of the coupling electric field is explored. We demonstrate that by appropriately adjusting the coupling electric field's amplitude or frequency, one can control the occurrence of two-photon resonances, providing additional sensitivity enhancement for the Rydberg sensor within the significantly extended off-resonance domain. The study underscores the significance of coupling fields in enhancing the sensitivity of Rydberg atom electric field sensors. These insights hold promising implications for the development of more robust and versatile electric field sensing devices, applicable in diverse fields such as precision measurements and quantum information processing.
\end{abstract}

\section{Introduction}
Rydberg atoms \cite{gallagher1988rydberg, gallagher1994rydberg, lim2013review} are characterized by highly excited states, where one or more electrons undergo a promotion to energy levels situated far from the atomic nucleus. These states derive their name from Johannes Rydberg, a Swedish physicist who extensively investigated atomic spectra during the late 19th century. Owing to their large principal quantum numbers, Rydberg atoms possess distinctive attributes including exceedingly vast atomic sizes \cite{fabre1983measuring} and extended lifetimes \cite{theodosiou1984lifetimes}.

The significance of Rydberg atoms in electric field sensing emerges from their pronounced sensitivity to electric fields. Even feeble electric fields can induce substantial alterations in the energy levels of Rydberg atoms, rendering them highly responsive to fluctuations in electric fields. The interplay between electric fields and Rydberg atoms can be effectively harnessed for a multitude of applications, with electric field sensing standing out as particularly prominent. By monitoring the shifts in energy levels or other quantifiable properties of Rydberg atoms, it becomes feasible to precisely measure and characterize electric fields. This capability bears crucial implications in various domains, including quantum information processing \cite{li2018unconventional, beterov2020application}, metro-logy \cite{de2000metrology, simons2021rydberg, ding2022enhanced}, and sensing technology \cite{artusio2022modern, liu2022highly}.

A key advantage of electric field sensing based on Rydberg atoms lies in its heightened sensitivity. The sizable dipole moments and prolonged lifetimes of Rydberg states enable exceedingly accurate measurements of electric fields, even at low magnitudes. Furthermore, this sensitivity can be further augmented by employing techniques such as electromagnetically induced transparency or coherent control of Rydberg atoms. Another noteworthy advantage is the versatility of sensors based on Rydberg atoms. They can be implemented across diverse environments, ranging from gaseous media to ultra-cold atomic ensembles or solid-state systems. This flexibility enables the adaptation of sensing techniques based on Rydberg atoms to a wide array of applications, including electric field mapping \cite{menendez2003stark, de2018accurate}, detection of weak signals \cite{grankin2014quantum, jia2021span}, and probing of electric field distributions in complex systems \cite{main1994rydberg, dunning2009engineering, holloway2018quantum}.

Some researchers has employed the one-mode Floquet theory \cite{shirley1965solution, dion1976time} to compute the sensitivity of such sensors, corroborating their findings through experiments \cite{meyer2020assessment}. Nevertheless, their results indicate a notable decrease in sensitivity when the signal frequency deviates from the resonance frequency, rendering this sensor incapable of providing accurate measurements in such instances. We have successfully addressed the issue theoretically by introducing an extra coupling electric field. Through the utilization of the many-mode Floquet theory \cite{ho1983semiclassical, ho1984semiclassical2, ho1985semiclassical3, ho1985semiclassical4}, we have calculated the sensitivity of the electric field sensor after the incorporation of the coupling electric field. By adjusting the amplitude or frequency of the coupling electric field, it becomes possible to achieve exceptionally high sensitivity for signal electric fields with frequency deviating from the resonance frequency.

In this paper, we investigate the sensitivity of Rydberg sensors using many-mode Floquet theory. We demonstrate how the coupling electric field influences the resonance peaks, leading to enhanced sensitivity in detecting electric fields. Through detailed analysis and numerical simulations, we explore the effects of different parameters, such as the amplitude and frequency of the coupling electric field, on the sensor's sensitivity. The findings of this study pave the way for designing and optimizing Rydberg atom electric field sensors for various applications \cite{saffman2010quantum, weimer2010rydberg, lee2012collective, saffman2016quantum, adams2019rydberg}.

\section{Background}\label{Background}
The prevailing approach to realize a Rydberg atom electric field sensor in practice typically encompasses several key steps. Firstly, optical pumping is employed to prepare the atoms in a specialized Rydberg state, which exhibits a heightened sensitivity to signal electric fields. Subsequently, the prepared state interacts with the signal electric field through Stark shifts, thereby inducing a collective phase shift. Finally, the collective phase shift is measured through optical readout, enabling the determination of the Rydberg state's response to the signal electric field.

The signal-to-noise ratio (SNR) inherent in this particular process is fundamentally constrained by the phase resolution exhibited by the readout stage, primarily stemming from two factors: the finite quantity of engaged Rydberg atoms and the standard quantum limit \cite{cox2018quantum}.
\subsection{Sensitivity of Rydberg atom electric field sensor}
We use the minimum detectable field (MDF) to assess the sensitivity of the Rydberg sensor. The MDF and the SNR are related concepts in the context of electric field sensing. The MDF refers to the weakest signal electric field that can be reliably sensed from noise in a system of Rydberg atoms. It represents the threshold below which the system's phase shift would be indistinguishable from noise. On the other hand, the SNR is a measure of the Stark shift relative to the background noise level. A higher SNR indicates a larger Stark shift, making it easier to be sensed.

As the SNR increases, the MDF decreases. When the Stark shift is larger relative to the noise, the system becomes more sensitive, and weaker signal electric field can be reliably detected. Conversely, as the SNR decreases, the MDF increases. When the noise level is high or comparable to the Stark shift, it becomes more challenging to distinguish the Stark shift from the noise, and a stronger signal electric field is required to achieve reliable detection.

Previous scholarly investigations on Rydberg atom electric field sensors have proposed a formulation for the SNR as follows
\begin{equation}
\text{SNR}=\frac{\varphi}{\Delta\varphi},
\end{equation}
where $\varphi=\Omega\tau$ represents the accumulated phase between two Rydberg states during an effective evolution time $\tau$. The Stark shift $\Omega(\epsilon_\text{s},\omega_\text{s})$ is a function of the amplitude $\epsilon_\text{s}$  and the frequency $\omega_\text{s}$ of the signal electric field. The phase noise $\Delta\varphi$ is assumed to adhere to the standard quantum limit (SQL), i.e. $\Delta\varphi=\Delta\varphi_{_\text{SQL}}=1/\sqrt{N}$, where $N$ denotes the number of Rydberg atoms.

To account for the finite coherence time $T$ of our Rydberg sensor, the effective evolution time $\tau$ depends on whether the measurement time $t$ is greater or less than $T$. When $t>T$, the optically-pumped Rydberg state collapses before readout. Consequently, the accumulated phase from an ensemble of atoms is reduced by a factor of $\sqrt{T/t}$. For the purposes of this study, we adopt a conservative value of $T=52\text{ns}$. The MDF is therefore defined as the amplitude $\epsilon_\text{s}$ of the signal electric field that yields $\text{SNR}=\sqrt{NT}\Omega(\epsilon_\text{s},\omega_\text{s})=1$ when measured over 1 second.

\subsection{One-mode Floquet theory}
In our investigation, we choose Rb Rydberg atoms and utilize the one-mode Floquet theory \cite{shirley1965solution} to compute the Stark shift $\Omega(\epsilon_\text{s},\omega_\text{s})$. Specifically, we focus on a single target state$\ket{50D_{5/2,m_J=1/2}}$ of Rb Rydberg atoms, and analyze approximately 40 Rydberg states that exhibit direct dipole-allowed transitions with the target state. These Rydberg states are $\ket{(50\pm\Delta n)P_{3/2}}$ and $\ket{(50\mp\Delta n)F_{7/2}}$, where$\Delta n=10$. We collectively denote all these states, including $\ket{50D_{5/2,m_J=1/2}}$, as $\{\ket{\alpha}\}$ and their corresponding energies as $\{E_\alpha\}$.

We approach the Rydberg atom from a quantum mechanical perspective while treating the signal electric field classically. Let us consider the Schrödinger equation

\begin{equation}\label{SE}
i\hbar\frac{\partial}{\partial t}\psi(t)=[H_0+V_\text{s}(t)]\psi(t),
\end{equation}
where $H_0$ represents the unperturbed Hamiltonian of the Rydberg atom, which possesses eigenstates $\{\ket{\alpha}\}$. Under the electric dipole approximation, the interaction Hamiltonian $V(t)$ between the signal electric field and the Rydberg atom can be expressed as

\begin{equation}
V(t)=-\boldsymbol{\mu}\cdot\boldsymbol{\epsilon}_\text{s}\cos(\omega_\text{s} t),
\end{equation}

where $\boldsymbol{\mu}$ represents the electric dipole moment operator, $\epsilon_\text{s}=|\boldsymbol{\epsilon}_\text{s}|$ and $\omega_\text{s}$ represent the amplitude and frequency of the signal electric field, respectively. In order to rewrite equation (\ref{SE}) in a matrix form, we can express it in the basis $\{\ket{\alpha}\}$. The accurate numerical values for the elements of $H_0$ and $V(t)$ in matrix form are determined through numerical integration, which is facilitated by the Alkali-Rydberg-Calculator (ARC) Python package.

Shirley suggested that we could select the basis $\{\ket{\alpha n}\}$ ($\ket{n}=e^{in\omega_\text{s} t} ,n\in\mathbb{Z}$) and convert the time-dependent Hamiltonian matrix in basis $\{\ket{\alpha}\}$ into a time-independent one-mode Floquet Hamiltonian matrix $H_\text{F}^1$
\begin{equation}\label{FH}
    \bra{\alpha n}H_\text{F}^1\ket{\beta m}=H_{\alpha\beta}^{n-m}+n\omega_\text{s}\delta_{\alpha\beta}\delta_{nm},
\end{equation}
where
\begin{equation}
    H_{\alpha\beta}^k=E_{\alpha}\delta_{\alpha\beta}\delta_{k,0}-\frac{1}{2}\bra{\alpha}\boldsymbol{\mu}\cdot\boldsymbol{\epsilon}\ket{\beta}(\delta_{k,1}+\delta_{k,-1}).
\end{equation}
All the eigenvalues of the Hamiltonian $H_\text{F}^1$ can be expressed as $\{E_\alpha^\prime+n\hbar\omega_\text{s}\}$. We can readily determine the corresponding energy $E^\prime_{50D}$ of the target Rydberg state $\ket{50D}$. Finally, we obtain the Stark shift
\begin{equation}
    \Omega(\epsilon_\text{s},\omega_\text{s}) = E^\prime_{50D}-E_{50D}.
\end{equation}

\begin{figure}
\centering
\includegraphics[width=0.7\textwidth]{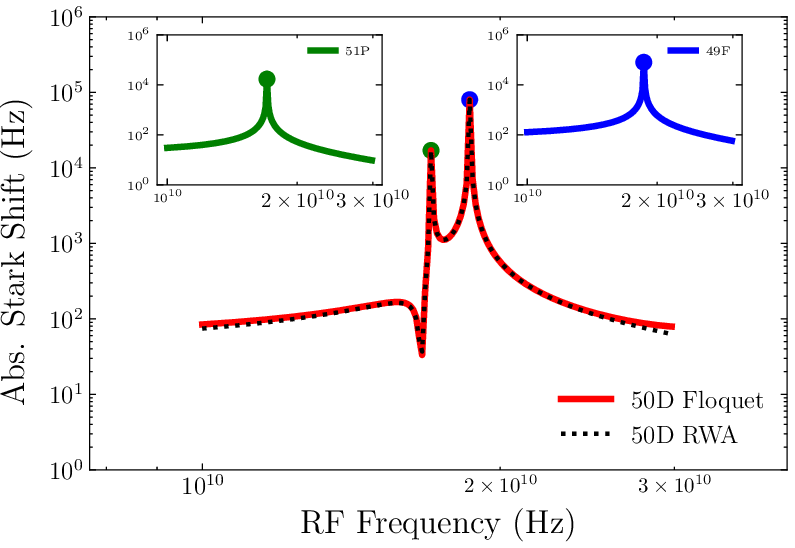}
\caption{\label{2}Absolute Stark shift as a function of RF frequency for the Rydberg state $\ket{50D}$. The red solid line represents the shift obtained using one-mode Floquet theory, while the black dotted line corresponds to the shift calculated using the rotating wave approximation (RWA). The green marker indicates the resonance between $\ket{50D}$ and $\ket{51P}$, while the blue marker represents the resonance between $\ket{50D}$ and $\ket{49F}$. Insets provide zoomed-in views of the Stark shift for $\ket{51P}$ and $\ket{49F}$. Logarithmic scales were used for all axes.}
\end{figure}

As we can see from figure \ref{2}, the outcome obtained through Floquet theory demonstrates a notable concurrence with the result derived from the rotating wave approximation (RWA). This concurrence provides compelling evidence attesting to the accuracy and reliability of the outcomes produced by the Floquet theory.

In the present investigation, we employ a total of 1,000 Rydberg atoms. The signal electric field amplitude $\epsilon_\text{s}$ was set at 0.1 V/m in order to evaluate the corresponding Stark shift $\Omega(0.1,\omega_\text{s})$. Another calculation was performed to determine the MDF, i.e. $\epsilon_{\text{s,min}}(\omega_\text{s})$ in a period of 1 second based on $\sqrt{NT}\Omega(\epsilon_\text{s},\omega_\text{s})=1$.

\begin{figure}
\centering
\includegraphics[width=0.7\textwidth]{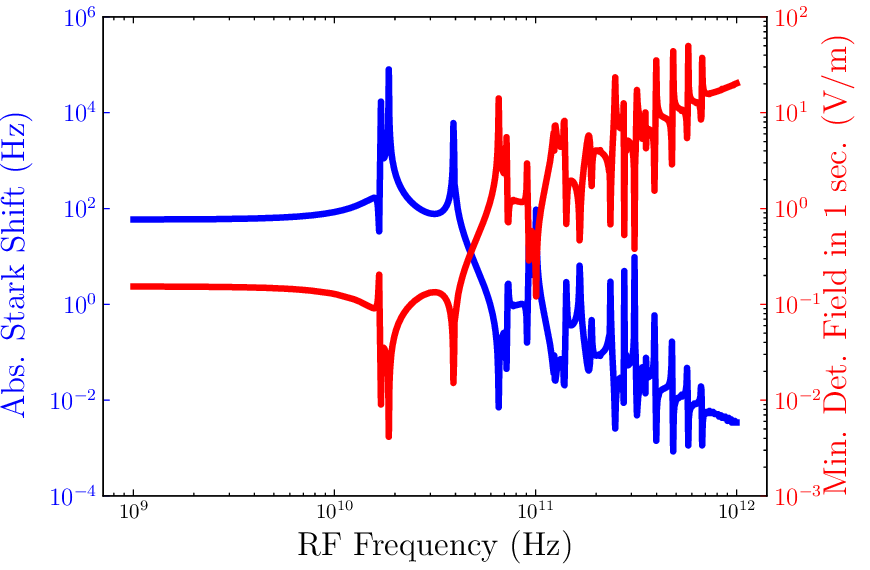}
\caption{\label{1}Absolute Stark shift for the Rydberg state $\ket{50D}$ and minimum detectable field in 1 second as functions of RF frequency. The blue line represents the absolute Stark shift, while the red line indicates the minimum detectable field in 1 second. Logarithmic scales were used for all axes.}
\end{figure}

Each peak of the blue line $\Omega(0.1,\omega_\text{s})$ in figure \ref{1} represents a resonance between $\ket{50D_{5/2,m_J=1/2}}$ and another Rydberg state under consideration. For instance, the peak at 17.03 GHz corresponds to the resonance between $\ket{51P_{3/2}}$ and $\ket{50D_{5/2,m_J=1/2}}$ while the peak at 18.63 GHz corresponds to the resonance between$\ket{49F_{7/2}}$ and $\ket{50D_{5/2,m_J=1/2}}$. The two peaks mentioned above appear close to each other because $\ket{50D_{5/2,m_J=1/2}}$ is positioned nearly symmetrically between $\ket{51P_{3/2}}$ and $\ket{49F_{7/2}}$. In a general sense, $\ket{50D_{5/2,m_J=1/2}}$ is nearly symmetrically positioned between any pair of $\ket{(50\pm\Delta n)P_{3/2}}$ and $\ket{(50\mp\Delta n)F_{7/2}}$. As a result, the remaining peaks in the blue line actually represent pairs of adjacent resonances.

The red line $\epsilon_{\text{s,min}}(\omega_\text{s})$ in figure \ref{1} exhibits apparent mirror symmetry with the blue line. This symmetry suggests the existence of a quantitative relationship between $\epsilon_{\text{s,min}}(\omega_\text{s})$ and $\Omega(0.1,\omega_\text{s})$. In the near resonance regime, the Stark shift $\Omega(\epsilon_\text{s},\omega_\text{s})\propto n^2\epsilon_\text{s}$ \cite{meyer2020assessment}, where $n$ denotes the principal quantum number of the Rydberg state. Given the definition of MDF as $\sqrt{NT}\Omega(\epsilon_\text{s},\omega_\text{s})=1$, we can readily deduce the following relationship
\begin{equation}
\epsilon_{\text{s,min}}(\omega_\text{s})\propto\frac{1}{\sqrt{NT}n^2}.
\end{equation}
In our investigation, we have set $\epsilon_\text{s}$=0.1 V/m and obtained $\Omega(0.1,\omega_\text{s})$. Consequently, we observe that
\begin{equation}\label{RES}
\epsilon_{\text{s,min}}(\omega_\text{s})\Omega(0.1,\omega_\text{s})\propto\frac{0.1n^2}{\sqrt{NT}n^2}=\frac{0.1}{\sqrt{NT}},
\end{equation}
which remains independent of the signal electric field's frequency $\omega_\text{s}$.

Due to the quantitative relationship between $\epsilon_{\text{s,min}}(\omega_\text{s})$ and $\Omega(0.1,\omega_\text{s})$, the subsequent sections of this paper directly investigate the Stark shift, instead of MDF, to characterize the sensitivity of Rydberg sensors. In practice, we adopt another amplitude of the signal electric field $\epsilon_\text{s}=0.01$ V/m.

\section{Rydberg Sensor with Coupling Field}\label{M}
A scrutiny of figure \ref{1} reveals that the Rydberg sensor demonstrates pronounced sensitivity when measuring a signal electric field with frequency equal to a resonance frequency $\hbar\omega_\text{s}=|E_\alpha-E_{50D}|$. Nevertheless, when the signal frequency $\omega_\text{s}$ deviates from the resonance frequency $|E_\alpha-E_{50D}|/\hbar$, the sensitivity diminishes rapidly. To tackle this challenge, we propose an innovative approach that incorporates an additional coupling electric field characterized by a substantially larger amplitude $\epsilon_\text{c}>10$ V/m compared to the signal electric field $\epsilon_\text{s}<0.1$ V/m. Moreover, we establish the frequency of the coupling electric field to be equal to the resonance frequency $\hbar\omega_\text{s}=|E_{49F}-E_{50D}|$.

The introduction of an additional coupling electric field leads to splittings of Rydberg states, such as $\ket{49F}$ and $\ket{51P}$, into two dressed states. Consequently, resonance peaks associated with these Rydberg states also split into multiple sub-peaks. This phenomenon is commonly referred to as Autler-Townes effect.

In instances where a Rydberg state lacks a direct dipole-allowed transition with another Rydberg state, the presence of a coupling electric field can render the resonance between these two states achievable through a process known as two-photon absorption. Consequently, distinct resonance peaks associated with this particular form of resonance become apparent.

Through the adjustment of the amplitude of the coupling electric field $\epsilon_\text{c}$, it becomes possible to effectively modulate the frequencies associated with the resonance peaks. When any of these frequencies aligns closely with the frequency of the signal electric field $\omega_\text{s}$, The sensitivity of the Rydberg sensor exhibits a noteworthy enhancement.

\subsection{Many-mode Floquet theory}
After the inclusion of a coupling electric field, the Schrödinger equation for the Rydberg sensor can be reformulated as

\begin{equation}\label{SE2}
i\hbar\frac{\partial}{\partial t}\psi(t)=[H_0+V_\text{c}(t)+V_\text{s}(t)]\psi(t),
\end{equation}
where $V_\text{c}(t)$ represents the interaction between the coupling electric field and the Rydberg atom,

\begin{equation}
V_\text{c}(t)=-\boldsymbol{\mu}\cdot\boldsymbol{\epsilon}_\text{c}\cos(\omega_\text{c} t),
\end{equation}
where $\epsilon_\text{c}=|\boldsymbol{\epsilon}_\text{c}|$ and $\omega_\text{c}$ correspond to the amplitude and frequency of the coupling electric field, respectively. The elements of $V_\text{c}(t)$ in matrix form can be accurately computed using the Alkali-Rydberg-Calculator (ARC) Python package.

We demonstrate the generalization of one-mode Floquet theory to encompass scenarios involving multiple electric fields \cite{ho1983semiclassical}. Without any loss of generality, we consider the case where two electric fields are involved—one being the signal electric field to be sensed and the other being the coupling electric field introduced in our Rydberg atoms.

To proceed, we adopt a strategy involving the introduction of a parameter $\omega$, which satisfies the following relationships

\begin{equation}\label{RE}
\omega_\text{c}=N_\text{c}\omega,\quad\omega_\text{s}=N_\text{s}\omega,
\end{equation}

where $N_\text{c}$ and $N_\text{s}$ represent integers that are relatively prime. It is important to note that the parameter $\omega$ only serves as an intermediate quantity and can be chosen to be arbitrarily small, ensuring that the equations \ref{RE} are satisfied with any desired level of precision. This flexibility is possible because, as we shall demonstrate later, the final result is found to be independent of the specific value of $\omega$.

Consequently, we are able to transform the Schrödinger equation \ref{SE2}, which involves two-mode incident electric fields with frequencies $\omega_\text{c}$ and $\omega_\text{s}$, into a Schrödinger equation with a one-mode incident electric field of frequency $\omega$. By employing the formula for the matrix elements of the one-mode Floquet Hamiltonian matrix (as given in equation \ref{FH}), we can derive the formula for the matrix elements of the two-mode Floquet Hamiltonian matrix $H_\text{F}^2$ in the basis ${\ket{\alpha n_\text{c} n_\text{s}}}$, where $\ket{n_\text{c}}=e^{in_\text{c}\omega_\text{c t}}$ and $\ket{n_\text{s}}=e^{in_\text{s}\omega_\text{s} t}$, with $n_\text{c}$ and $n_\text{s}$ being integers.
\begin{align}
    \bra{\alpha, n_\text{c}, n_\text{s}}H_\text{F}^2\ket{\beta, m_\text{c}, m_\text{s}}&=\bra{\alpha, n_\text{c}N_\text{c}+ n_\text{s}N_\text{s}}H_\text{F}^1\ket{\beta, m_\text{c}N_\text{c}+ m_\text{s}N_\text{s}}\notag\\
    &=H_{\alpha\beta}^{n_\text{c}-m_\text{c},n_\text{s}-m_\text{s}}+(n_\text{c}N_\text{c}+ n_\text{s}N_\text{s})\omega\delta_{\alpha\beta}\delta_{n_\text{c}m_\text{c}}\delta_{n_\text{s}m_\text{s}}\notag\\
    &=H_{\alpha\beta}^{n_\text{c}-m_\text{c},n_\text{s}-m_\text{s}}+(n_\text{c}\omega_\text{c}+ n_\text{s}\omega_\text{s})\delta_{\alpha\beta}\delta_{n_\text{c}m_\text{c}}\delta_{n_\text{s}m_\text{s}},
\end{align}
where $H_{\alpha\beta}^{k_\text{c}, k_\text{s}}$ is defined as

\begin{equation}
    H_{\alpha\beta}^{k_\text{c}, k_\text{s}}=E_{\alpha}\delta_{\alpha\beta}\delta_{k_\text{c},0}\delta_{k_\text{s},0}-\frac{1}{2}\bra{\alpha}\boldsymbol{\mu}\cdot\boldsymbol{\epsilon}_\text{c}\ket{\beta}(\delta_{k_\text{c},1}+\delta_{k_\text{c},-1})-\frac{1}{2}\bra{\alpha}\boldsymbol{\mu}\cdot\boldsymbol{\epsilon}_\text{s}\ket{\beta}(\delta_{k_\text{s},1}+\delta_{k_\text{s},-1}).
\end{equation}

The eigenvalues of the Hamiltonian $H_\text{F}^2$ can be expressed as $\{E_\alpha^{\prime\prime}+\hbar n_\text{c}\omega_\text{c}+\hbar n_\text{s}\omega_\text{s}\}$. From these eigenvalues, we can readily determine the corresponding energy $E^{\prime\prime}_\alpha$ and the calculate the Stark shift $\Omega_{\alpha}(\epsilon_\text{s},\omega_\text{s})$ for any Rydberg states under consideration.

\subsection{Autler-Townes effect}
\begin{figure}
\centering
\includegraphics[width=0.7\textwidth]{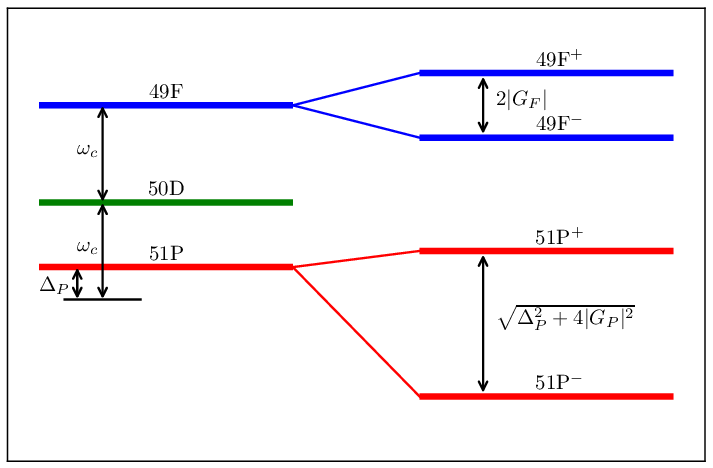}
\caption{\label{3}Energy level diagram depicting the splittings of Rydberg states $\ket{49F}$ and $\ket{51P}$ under the influence of a coupling electric field with frequency matching the resonance between Rydberg states $\ket{50D}$ and $\ket{49F}$.}
\end{figure}

In the resonance or near-resonance regime, the Autler-Townes effect causes the splittings of Rydberg states into two distinct dressed states. As illustrated in Figure \ref{3}, the Rydberg state $\ket{49F}$ undergoes a split and gives rise to two dressed states $\ket{49F^+}$ and $\ket{49F^-}$. The Rydberg state $\ket{51P}$ experiences a similar splitting, resulting in the formation of two dressed states $\ket{51P^+}$ and $\ket{51P^-}$. Apply rotating wave approximation (RWA) to the Hamiltonian of a reduced two-level system and solve the corresponding eigenvalue equation $H_\alpha\ket{\pm}_\alpha=\lambda_\alpha^\pm\ket{\pm}_\alpha$, we can derive an approximate expression for the eigenvalues of the dressed states
\begin{equation}\label{RWA}
\lambda_\alpha^\pm = \frac{1}{2}(\Delta_\alpha \pm \sqrt{\Delta_\alpha^2 + 4|G_\alpha|^2}), \end{equation}
where $\alpha$ takes on the values of either $F$ or $P$, which correspond to the Rydberg states $\ket{49F}$ or $\ket{51P}$, respectively. $\Delta_\alpha$ represents the detuning of the coupling electric field from the resonance between the Rydberg states $\ket{\alpha}$ and the target state $\ket{50D}$. We define the coupling coefficients $G_\alpha=\mu_\alpha\epsilon_\text{c}/\hbar$, where $\mu_\alpha$ is the dipole moment of the transition between $\ket{50D}$ and $\ket{\alpha}$ while $\epsilon_\text{c}$ is the amplitude of the coupling electric field.

In the considered parameter regimes, only interactions among specific Rydberg states ($\ket{49F}$, $\ket{50D}$, $\ket{51P}$) and particular dressed states ($\ket{49F^\pm}$, $\ket{51P^\pm}$) emerge. To enhance clarity, we denote these states as $\ket{F}$, $\ket{D}$, $\ket{P}$, $\ket{F^\pm}$, and $\ket{P^\pm}$, respectively. Additionally, we represent the resonances between these states as $(\alpha,\beta)$, where $\alpha$ and $\beta$ could be any of the aforementioned states. For instance, $(D,F)$ signifies the resonance between the Rydberg states $\ket{50D}$ and $\ket{49F}$, and $(D,P^+)$ represents the resonance between the Rydberg state $\ket{50D}$ and the dressed state $\ket{51P^+}$.

\begin{figure}
  \centering
  \subfigure[]{
  \label{44}
  \includegraphics[width=0.4\textwidth]{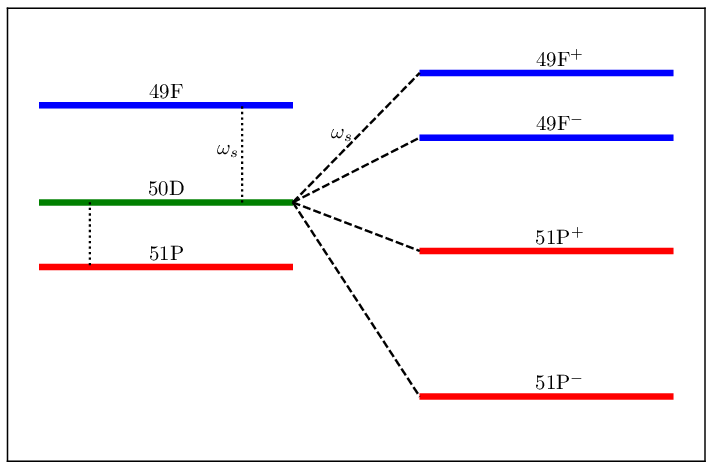}
  }
  \subfigure[]{
  \label{41}
  \includegraphics[width=0.4\textwidth]{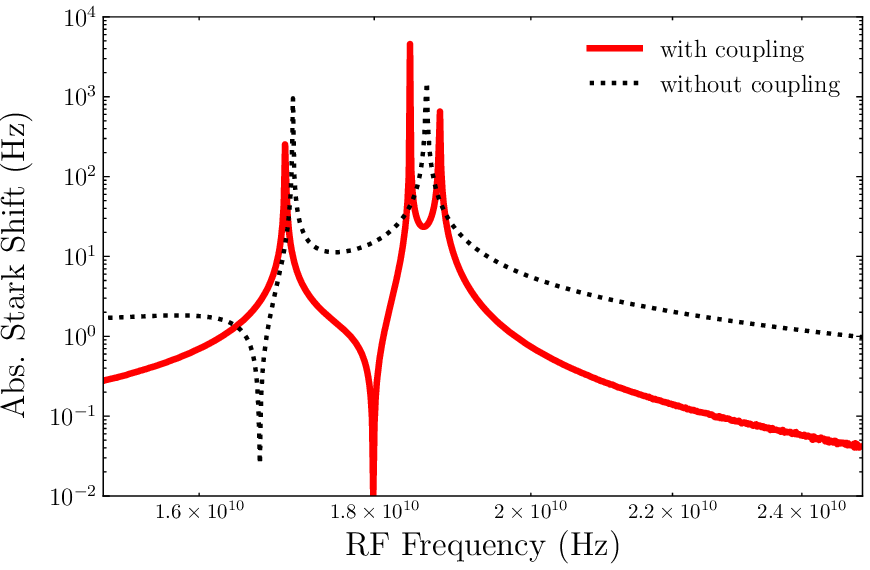}
  }
  \subfigure[]{
  \label{42}
  \includegraphics[width=0.4\textwidth]{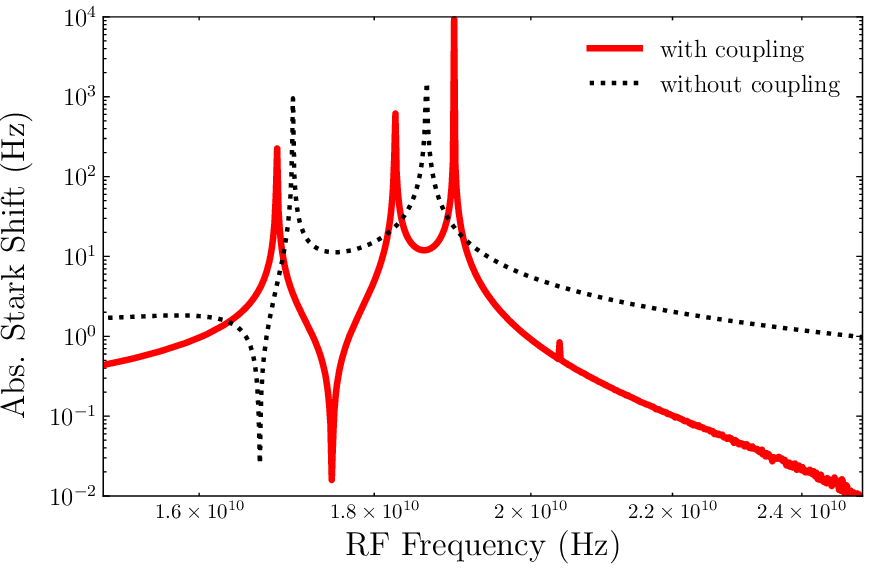}
  }
  \subfigure[]{
  \label{43}
  \includegraphics[width=0.4\textwidth]{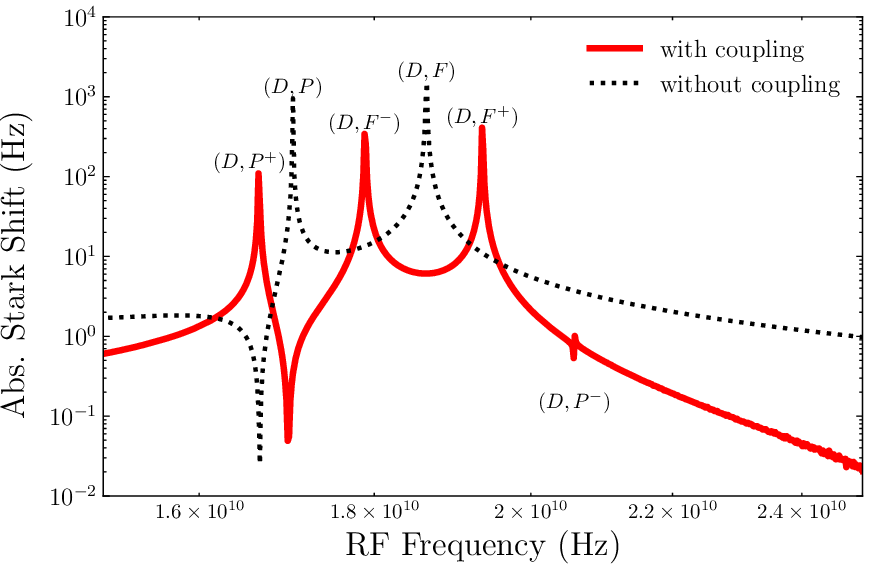}
  }
  \caption{(a) Energy level diagram depicting the resonance before and after using a coupling electric field. The dotted lines represent the two kinds of resonance before using a coupling electric field, while the dashed lines represent the four kinds of resonance after using a coupling electric field. (b)(c)(d) Comparison of absolute Stark shift as a function of RF frequency for the Rydberg state $\ket{50D}$ with and without coupling electric field. The amplitude of the coupling electric field is 10, 20, and 40 V/m, respectively. The red line represents the absolute Stark shift after using a coupling electric field, while the black dotted line represents that before using a coupling electric field. The absolute Stark shift is on a logarithmic scale.}
  \label{4}
\end{figure}

Due to the typically small amplitude of the signal electric field ($\epsilon_\text{s}<0.1$V/m), the Autler-Townes effect can be disregarded in the absence of a coupling electric field. The dotted line in figure \ref{43} exhibits two resonance peaks. The resonance peak occurring at 17.04 GHz is associated with the resonance between the Rydberg states $\ket{D}$ and $\ket{P}$, represented by $(D,P)$, while the resonance peak at 18.65 GHz corresponds to the resonance between $\ket{D}$ and $\ket{F}$, donated as $(D,F)$. The dotted lines in figure \ref{44} indicate the two kinds of resonance mentioned above. It is noteworthy that the Rydberg states do not split into two dressed states in the absence of a coupling electric field.

In contrast, when a strong coupling electric field is introduced ($\epsilon_\text{c}>10$ V/m), the Autler-Townes effect becomes evident. As we can see from figures \ref{41}, \ref{42}, and \ref{43}, the resonance peaks in dotted lines split into two sub-peaks in red lines, respectively. To explain this phenomenon, we select a coupling electric field with a frequency equal to the resonance frequency between the Rydberg states $\ket{D}$ and $\ket{F}$, i.e. $\hbar\omega_\text{c}=|E_{F}-E_{D}|$. The chosen coupling electric field possesses amplitudes ($\epsilon_\text{c}$=10, 20 and 40 V/m) significantly greater than that of the signal electric field.

Upon examination of the Rydberg state $\ket{F}$, the detuning $\Delta_F$ becomes zero, and the eigenvalues of the dressed states $\ket{F^\pm}$
\begin{equation}
    \lambda_F^\pm=\pm|G_F|.
\end{equation}
Notably, the Rydberg state $\ket{F}$ exhibits a symmetrical splitting into two dressed states, $\ket{F^+}$ and $\ket{F^-}$, each possessing energy levels of $E_{F}+|G_F|$ and $E_{F}-|G_F|$, respectively.

As depicted in figures \ref{41}, \ref{42}, and \ref{43}, the resonance peaks corresponding to the Rydberg state $\ket{F}$ also undergo a symmetrical splitting and result in two sub-peaks. Consider the red line depicted in figure \ref{43} as an illustrative instance. It is noteworthy that the peak observed at 17.88 GHz corresponds to the resonance between Rydberg state $\ket{D}$ and dressed state $\ket{F^-}$, represented by $(D,F^-)$, while the peak observed at 19.35 GHz is associated with the resonance between Rydberg state $\ket{D}$ and dressed state $\ket{F^+}$, donated as $(D,F^+)$. As we progressively enhance the amplitude $\epsilon_\text{c}$ of the coupling electric field from 10 V/m to 40 V/m, the energy gap $E_{F}^+-E_{F}^-=2|G_F|$ between these two dressed states $\ket{F^+}$ and $\ket{F^-}$ becomes larger. Therefore, as we can see from figures \ref{41}, \ref{42}, and \ref{43}, the two sub-peaks $(D,F^+)$ and $(D,F^-)$ become increasingly distant from the resonance peak $(D,F)$. This behavior arises due to the increased coupling coefficients $|G_F|$ associated with the amplification of $\epsilon_\text{c}$.

Let us now redirect our attention to the discussion of the Rydberg state $\ket{P}$. Because the detuning of the coupling electric field from the resonance between the Rydberg states $\ket{D}$ and $\ket{P}$ is exceedingly small, we can employ equation (\ref{RWA}) to provide a rough estimation of the eigenvalues of the dressed states $\ket{P^\pm}$
\begin{equation}
\lambda_P^\pm = \frac{1}{2}(\Delta_P \pm \sqrt{\Delta_P^2 + 4|G_P|^2}).
\end{equation}

For illustrative purposes, we shall consider the red line depicted in figure \ref{43}. The presence of a short peak at 20.60 GHz indicates the resonance between the Rydberg state $\ket{D}$ and the dressed state $\ket{P^-}$, while the tall peak at 16. GHz signifies the resonance between the Rydberg state $\ket{D}$ and the dressed state $\ket{P^+}$. The Rydberg state $\ket{P}$ no longer exhibits a symmetrical splitting into two dressed states. As a result, the resonance peaks associated with the Rydberg state $\ket{P}$ no longer display a symmetrical splitting like $\ket{F}$. As we can see from figure \ref{41}, the peak corresponding to the resonance between the Rydberg state $\ket{D}$ and the dressed state $\ket{P^-}$ is invisible when the amplitude $\epsilon_\text{c}$ of the coupling electric field is 10 V/m. However, as we amplify $\epsilon_\text{c}$ from 10 V/m to 40 V/m, the peak grows increasingly evident. In addition, when we progressively intensify the amplitude $\epsilon_\text{c}$ of the coupling electric field, the energy gap $E_{P}^+-E_{P}^-=\sqrt{\Delta_P^2+4|G_P|^2}$ between the two dressed states mentioned above also exhibits an increase. This outcome can be attributed to the augmented coupling coefficients $|G_P|$ resulting from the enhanced amplitude $\epsilon_\text{c}$.

The splitting of resonance peaks holds immense importance in the realm of electric field sensing. As an illustration, let us examine a signal electric field with a frequency of 18.99 GHz, which is close to but not precisely equal to the resonance frequency between the Rydberg states $\ket{F}$ and $\ket{D}$. Without any coupling electric field, the Stark shift caused by the signal electric field is 23.21 Hz, as we can see from the dotted line in figure \ref{42}. This implies that the minimum detectable (MDF) field at 18.99 GHz is approximately 0.53 V/m. However, figure \ref{42} provides a clear demonstration that when a coupling electric field with an amplitude of 20 V/m is applied, it leads to a noteworthy increase in the Stark shift caused by the same signal electric field, amounting to 9.32 kHz. Consequently, the MDF at the same frequency decrease to about 1.07 mV/m. In general, by tuning the amplitude $\epsilon_\text{c}$ of the coupling electric field, it becomes feasible to efficiently control the frequencies corresponding to the resonance peaks. When any of these frequencies matches the frequency $\omega_\text{s}$ of the signal electric field, the Rydberg sensor's sensitivity experiences a remarkable improvement.

\subsection{Two-photon absorption}
Introducing a coupling electric field not only leads to the splitting of original resonance peaks but also gives rise to additional resonance peaks through two-photon absorption. For the sake of clarity, we shall refer to this kind of resonance as ``two-photon resonance".

Prior to the incorporation of the coupling electric field, the absence of direct dipole-allowed transition hindered the occurrence of resonance between certain Rydberg states. Nevertheless, with the inclusion of the coupling electric field, the possibility of two-photon resonance emerges. Through the manipulation of the coupling electric field's amplitude $\epsilon_\text{c}$ and frequency $\omega_\text{c}$, we gain the capability to regulate the frequency at which the two-photon resonance takes place.

This capability is of significant importance in the field of electric field sensing. The presence of two-photon resonance peaks allows us to detect and analyze electric fields more effectively. When the frequency corresponding to the two-photon resonance peak aligns with the signal electric field's frequency $\omega_\text{s}$, the Rydberg sensor exhibits a notable enhancement in sensitivity.

We shall initiate our investigation to the phenomenon of two-photon absorption in the resonance regime $\hbar\omega_\text{c}=|E_{F}-E_{D}|$. Within this particular regime, a notable effect known as the Autler-Townes effect manifests, leading to the splitting of the Rydberg state into a pair of dressed states. Specifically, the initial Rydberg state $\ket{F}$ undergoes this splitting, yielding two distinct dressed states $\ket{F^+}$ and $\ket{F^-}$. A similar outcome is observed for the Rydberg state $\ket{P}$, which likewise experiences a splitting, giving rise to the two dressed states $\ket{P^+}$ and $\ket{P^-}$. It is worth noting that the phenomenon of two-photon resonance can be discerned between certain pairs of these dressed states.

\begin{figure}
  \centering
    \subfigure[]{
  \label{54}
  \includegraphics[width=0.4\textwidth]{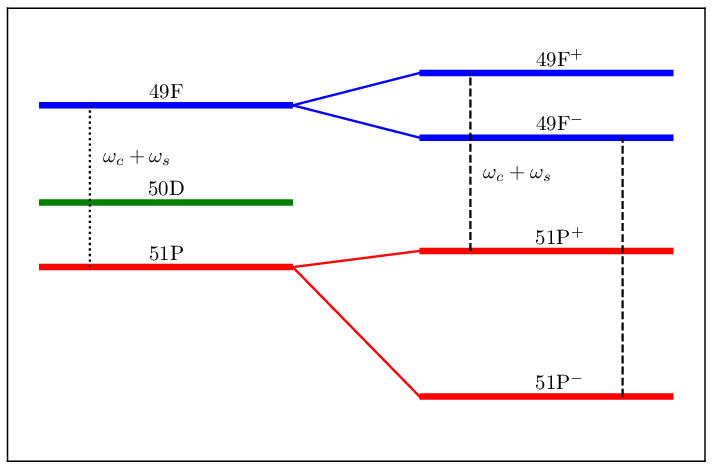}
  }
    \subfigure[]{
  \label{51}
  \includegraphics[width=0.4\textwidth]{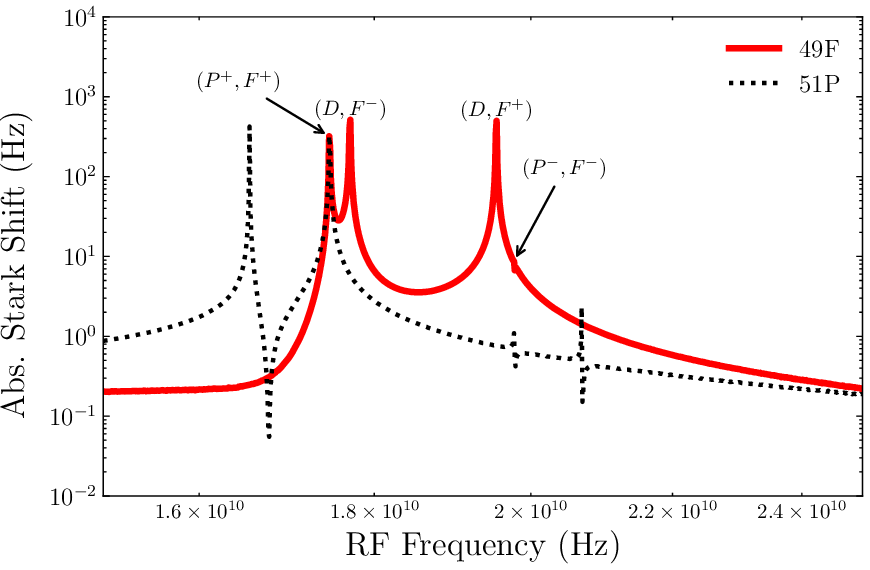}
  }
  \subfigure[]{
  \label{52}
  \includegraphics[width=0.4\textwidth]{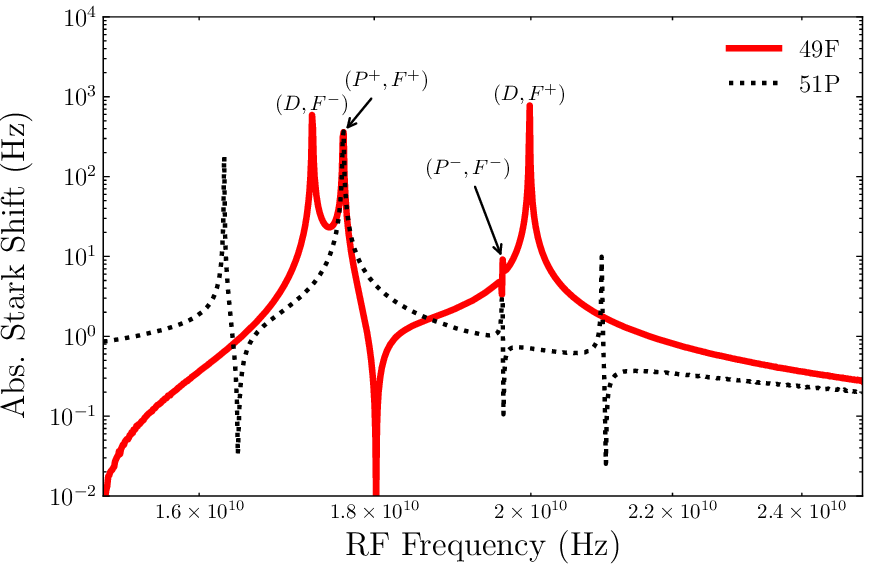}
  }
  \subfigure[]{
  \label{53}
  \includegraphics[width=0.4\textwidth]{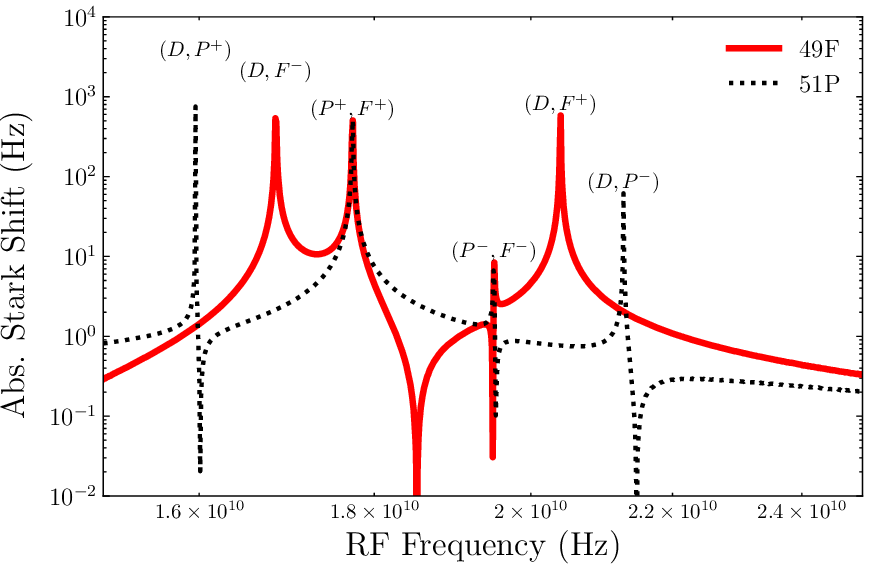}
  }
  \subfigure[]{
  \label{61}
  \includegraphics[width=0.4\textwidth]{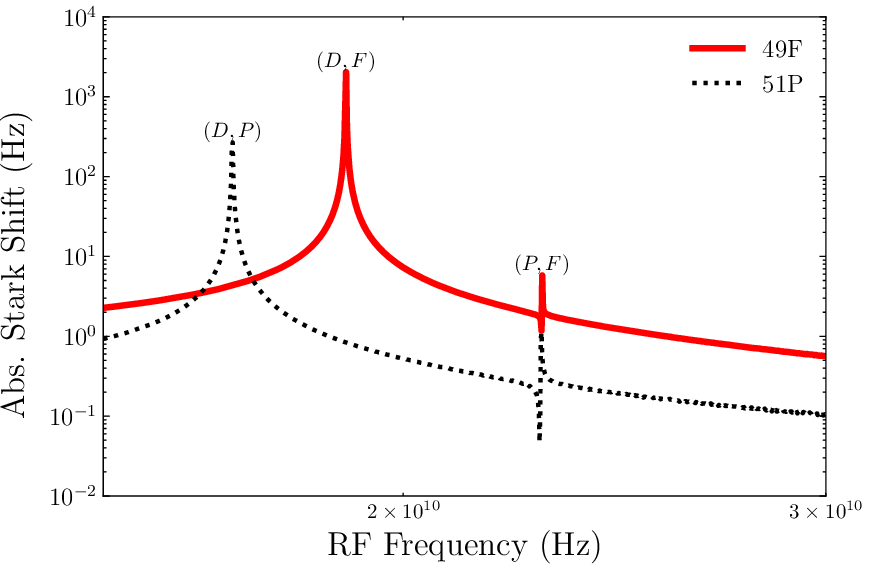}
  }
  \subfigure[]{
  \label{62}
  \includegraphics[width=0.4\textwidth]{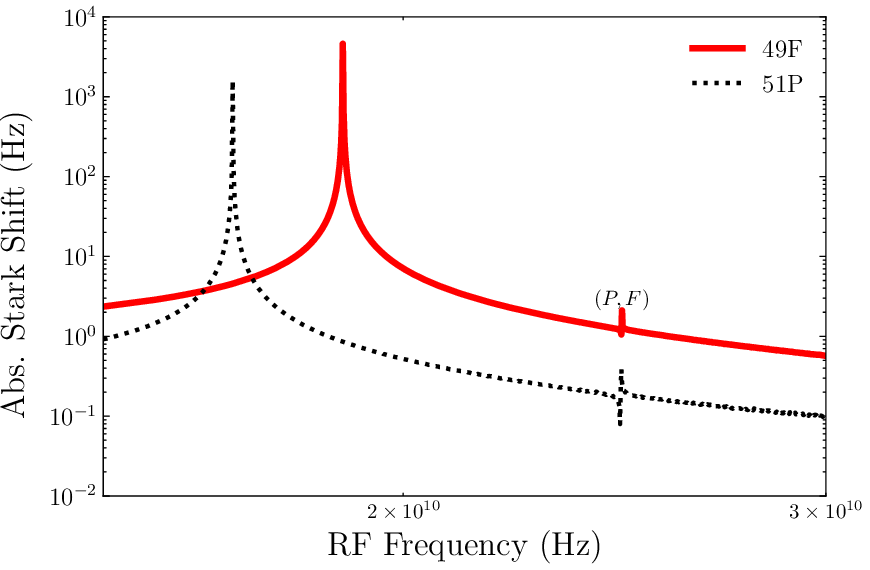}
  }
  \caption{(a) Energy level diagram depicting the two-photon absorption in the resonance and off-resonance regime. The dotted line represents the only kind of two-photon absorption in the off-resonance regime, while the dashed lines represent the two kinds of two-photon absorption in the resonance regime. (b)(c)(d) Comparison of absolute Stark shift as a function of RF frequency for the Rydberg state $\ket{49F}$ and $\ket{51P}$ in the resonance regime. The amplitude of the coupling electric field is 50, 75, and 100 V/m, respectively. The red line represents the absolute Stark shift for the Rydberg state $\ket{49F}$, while the black dotted line represents that for the Rydberg state $\ket{51P}$. The absolute Stark shift is on a logarithmic scale. (e)(f) Comparison of absolute Stark shift as a function of RF frequency for the Rydberg state $\ket{49F}$ and $\ket{51P}$ in the off-resonance regime. The amplitude of the coupling electric field is 100 V/m, while the frequency of the field is 0.7 and 0.6$(E_{49F}-E_{50D})/\hbar$, respectively. The absolute Stark shift is on a logarithmic scale.}
  \label{5}
\end{figure}

Figure \ref{51}, \ref{52}, or \ref{53} exhibits a total of eight resonance peaks, with four appearing on the red line and the other four on the dotted line. Remarkably, two of the peaks observed on the red line precisely coincide with two peaks on the dotted line, indicating two-photon resonances resulting from two-photon absorption. The remaining peaks are sub-peaks associated with resonances between the target Rydberg state $\ket{D}$ and the dressed states $\ket{P^+}$, $\ket{P^-}$, $\ket{F^+}$, and $\ket{F^-}$, as discussed previously.

Consider figure \ref{53} as a case in point. The observation of a pronounced peak at 17.74 GHz indicates the occurrence of two-photon resonance between the dressed states $\ket{P^+}$ and $\ket{F^+}$. For the purpose of enhancing clarity, the notation $(P^+,F^+)$ is adopted in figure \ref{53} to represent this two-photon resonance. Similarly, another peak at 19.50 GHz corresponds to the two-photon resonance between $\ket{P^-}$ and $\ket{F^-}$, donated by $(P^-,F^-)$ in figure \ref{53}. These two-photon resonances are visually represented by the dashed lines in figure \ref{54}.

The energy gap between the dressed states $\ket{P^+}$ and $\ket{F^+}$ can be mathematically expressed as
\begin{equation}\label{twop}
E_{F}^+ - E_{P}^+ = E_{F} - E_{P} + |G_F| - \frac{1}{2} \left( \Delta_P + \sqrt{\Delta_P^2 + 4|G_P|^2} \right) = \hbar(\omega_\text{c} + \omega_\text{s}).
\end{equation}
Similarly, for the dressed states $\ket{P^-}$ and $\ket{F^-}$, the energy gap is
\begin{equation}\label{twom}
E_{F}^- - E_{P}^- = E_{F} - E_{P} - |G_F| + \frac{1}{2} \left( \Delta_P + \sqrt{\Delta_P^2 + 4|G_P|^2} \right) = \hbar(\omega_\text{c} + \omega_\text{s}).
\end{equation}
It is established that $\hbar\omega_\text{c} = E_{F} - E_{D}$, allowing for the rapid determination of the frequency at which two-photon resonance occurs.

Figure \ref{51} illustrates that the two-photon resonance peak at 17.5 GHz, associated with the resonance between $\ket{P^+}$ and $\ket{F^+}$, is on the left side of the resonance peak at 17.7 GHz, corresponding to the resonance between $\ket{D}$ and $\ket{F^-}$. Subsequently, in figure \ref{52}, as the amplitude $\epsilon_\text{c}$ of the coupling electric field is amplified from  V/m to 75 V/m, the two-photon resonance peak undergoes a rightward shift relative to the resonance peak associated with the Rydberg state $\ket{D}$ and dressed state $\ket{F^-}$, donated by $(D,F^-)$ in figure \ref{52}. Specifically, using a coupling electric field with a amplitude of 75 V/m, the two-photon resonance peak $(P^+,F^+)$ occurs at 17.6 GHz, while the resonance peak $(D,F^-)$ is observed at 17.3 GHz. The two-photon resonance peak $(P^+,F^+)$ is now on the right side of the resonance peak $(D,F^-)$. Further amplification of $\epsilon_\text{c}$ in figure \ref{53} results in a further increase of the energy gap between the dressed states $\ket{P^+}$ and $\ket{F^+}$. As a result, the two-photon resonance peak $(P^+,F^+)$ at 17.7 GHz becoming increasingly distant from the resonance peak $(D,F^+)$ at 16.8 GHz.

Distinctively, the two-photon resonance peak corresponding to the dressed states $\ket{P^-}$ and $\ket{F^-}$, represented by $(P^-,F^-)$, exhibits a leftward shift relative to the resonance peak associated with the Rydberg state $\ket{D}$ and the dressed state $\ket{F^+}$, donated as $(D,F^+)$. This shift becomes apparent as the amplitude of the coupling electric field increases successively from  V/m to 100 V/m. The frequency corresponding to the two-photon resonance peak experiences a successive decrease from 20.7 GHz to 19.5 GHz. That is because the dipole moment $\mu_F$ for the transition between Rydberg states $\ket{D}$ and $\ket{F}$ exceeds the dipole moment $\mu_P$ for the transition between Rydberg states $\ket{D}$ and $\ket{P}$. Therefore, when we amplify the amplitude of the coupling electric field, the corresponding coupling coefficients $G_F$ experience a more rapid increase compared to $G_P$. As a result, from equation (\ref{twop}) and equation (\ref{twom}) we can find that when we amplify the amplitude of the coupling electric field, the energy gap between the dressed states $E_{F}^+$ and $E_{P}^+$ increases, while the energy gap between the dressed states $E_{F}^- - E_{P}^-$ decreases. That is why the two-photon resonance peak associated with the dressed states $\ket{P^+}$ and $\ket{F^+}$ experiences a rightward shift, while the two-photon resonance peak corresponding to the dressed states $\ket{P^-}$ and $\ket{F^-}$ undergoes a leftward shift.

In the off-resonance regime, the Autler-Townes effect becomes unobservable. However, the shift of the two-photon resonance peak is achievable by adjusting the frequency $\omega_\text{c}$ of the coupling electric field. Figure \ref{61} or \ref{62} demonstrates the presence of a total of four resonance peaks, instead of eight, along the red and dotted lines. Specifically, a single peak on the red line precisely aligns with a corresponding peak on the dotted line, representing two-photon resonance between the Rydberg states $\ket{F}$ and $\ket{P}$. Additionally, the other resonance peak on the red line corresponds to the resonance between Rydberg states $\ket{D}$ and $\ket{F}$, while the remaining resonance peak on the dotted line corresponds to the resonance between Rydberg states $\ket{D}$ and $\ket{P}$.

It is important to note that in the off-resonance regime, no Rydberg state splittings occur. As the frequency of the coupling electric field $\omega_\text{c}$ decreases from figure \ref{61} to figure \ref{62}, the frequency at which two-photon resonance takes place increases, following the relationship $\hbar(\omega_\text{c}+\omega_\text{s})=E_{F}-E_{P}$.

The appearance of two-photon resonance peaks bears significant importance in the domain of electric field sensing. Through appropriate adjustments of the amplitude $\epsilon_\text{c}$ or the frequency $\omega_\text{c}$ of the coupling electric field, precise control over the frequencies associated with the two-photon resonance peaks can be achieved. Upon the alignment of any of these frequencies with the frequency $\omega_\text{s}$ of the signal electric field, the sensitivity of the Rydberg sensor undergoes further enhancement.

\section{Conclusion}
In this paper, we have investigated the sensitivity of a Rydberg atom electric field sensor, focusing on the use of minimum detectable field (MDF) as a measure of sensitivity. We employed one-mode Floquet theory to compute the Stark shift for some Rydberg states in the presence of a signal electric field and demonstrated the concurrence of the results with those obtained using the rotating wave approximation (RWA).

To improve the sensitivity of the Rydberg sensor when the signal frequency deviates from resonance frequencies between Rydberg states, we incorporate an extra coupling electric field and use many-mode Floquet theory to theoretically analyze this kind of Rydberg sensor. We discussed the Autler-Townes effect resulting from the coupling field, which led to the splitting of Rydberg states into dressed states, and demonstrated how this effect enhances the sensitivity by modulating the frequencies of resonance peaks.

Furthermore, we explored the phenomenon of two-photon absorption in the presence of the coupling electric field. We showed that by adjusting the parameters of the coupling field, it is possible to effectively control the occurrence of two-photon resonances, providing an additional enhancement to the sensitivity of the Rydberg sensor within the significantly extended off-resonance domain.

Overall, our study highlights the importance of coupling fields in improving the sensitivity of Rydberg atom electric field sensors. These findings pave the way for the development of more robust and versatile electric field sensing devices that can be used in various applications, ranging from precision measurements to quantum information processing.

\newpage
\bibliographystyle{unsrt}
\bibliography{Rydberg}

\begin{thebibliography}{10}

\bibitem{gallagher1988rydberg}
Thomas~F Gallagher.
\newblock Rydberg atoms.
\newblock {\em Reports on Progress in Physics}, 51(2):143, 1988.

\bibitem{gallagher1994rydberg}
Thomas~F Gallagher.
\newblock Rydberg atoms.
\newblock In {\em Springer Handbook of Atomic, Molecular, and Optical Physics},
  pages 231--240. Springer, 1994.

\bibitem{lim2013review}
Jongseok Lim, Han-gyeol Lee, and Jaewook Ahn.
\newblock Review of cold rydberg atoms and their applications.
\newblock {\em Journal of the Korean Physical Society}, 63:867--876, 2013.

\bibitem{fabre1983measuring}
C~Fabre, M~Gross, JM~Raimond, and S~Haroche.
\newblock Measuring atomic dimensions by transmission of rydberg atoms through
  micrometre size slits.
\newblock {\em Journal of Physics B: Atomic and Molecular Physics},
  16(21):L671, 1983.

\bibitem{theodosiou1984lifetimes}
Constantine~E Theodosiou.
\newblock Lifetimes of alkali-metal-atom rydberg states.
\newblock {\em Physical Review A}, 30(6):2881, 1984.

\bibitem{li2018unconventional}
DX~Li and XQ~Shao.
\newblock Unconventional rydberg pumping and applications in quantum
  information processing.
\newblock {\em Physical Review A}, 98(6):062338, 2018.

\bibitem{beterov2020application}
II~Beterov, DB~Tretyakov, VM~Entin, EA~Yakshina, II~Ryabtsev, M~Saffman, and
  S~Bergamini.
\newblock Application of adiabatic passage in rydberg atomic ensembles for
  quantum information processing.
\newblock {\em Journal of Physics B: Atomic, Molecular and Optical Physics},
  53(18):182001, 2020.

\bibitem{de2000metrology}
B~De~Beauvoir, Catherine Schwob, O~Acef, L~Jozefowski, L~Hilico, F~Nez,
  L~Julien, A~Clairon, and F~Biraben.
\newblock Metrology of the hydrogen and deuterium atoms: Determination of the
  rydberg constant and lamb shifts.
\newblock {\em The European Physical Journal D-Atomic, Molecular, Optical and
  Plasma Physics}, 12(1):61--93, 2000.

\bibitem{simons2021rydberg}
Matthew~T Simons, Alexandra~B Artusio-Glimpse, Amy~K Robinson, Nikunjkumar
  Prajapati, and Christopher~L Holloway.
\newblock Rydberg atom-based sensors for radio-frequency electric field
  metrology, sensing, and communications.
\newblock {\em Measurement: Sensors}, 18:100273, 2021.

\bibitem{ding2022enhanced}
Dong-Sheng Ding, Zong-Kai Liu, Bao-Sen Shi, Guang-Can Guo, Klaus M{\o}lmer, and
  Charles~S Adams.
\newblock Enhanced metrology at the critical point of a many-body rydberg
  atomic system.
\newblock {\em Nature Physics}, 18(12):1447--1452, 2022.

\bibitem{artusio2022modern}
Alexandra Artusio-Glimpse, Matthew~T Simons, Nikunjkumar Prajapati, and
  Christopher~L Holloway.
\newblock Modern rf measurements with hot atoms: A technology review of rydberg
  atom-based radio frequency field sensors.
\newblock {\em IEEE Microwave Magazine}, 23(5):44--56, 2022.

\bibitem{liu2022highly}
Bang Liu, Li-Hua Zhang, Zong-Kai Liu, Zheng-Yuan Zhang, Zhi-Han Zhu, Wei Gao,
  Guang-Can Guo, Dong-Sheng Ding, and Bao-Sen Shi.
\newblock Highly sensitive measurement of a megahertz rf electric field with a
  rydberg-atom sensor.
\newblock {\em Physical Review Applied}, 18(1):014045, 2022.

\bibitem{menendez2003stark}
JM~Men{\'e}ndez, I~Mart{\i}n, and AM~Velasco.
\newblock Stark maps and rydberg transitions in the presence of an electric
  field for li, na, and k. a quantum defect orbital approach.
\newblock {\em The Journal of chemical physics}, 119(24):12926--12930, 2003.

\bibitem{de2018accurate}
Sylvain De~L{\'e}s{\'e}leuc, Sebastian Weber, Vincent Lienhard, Daniel Barredo,
  Hans~Peter B{\"u}chler, Thierry Lahaye, and Antoine Browaeys.
\newblock Accurate mapping of multilevel rydberg atoms on interacting spin-1/2
  particles for the quantum simulation of ising models.
\newblock {\em Physical review letters}, 120(11):113602, 2018.

\bibitem{grankin2014quantum}
Andrey Grankin, E~Brion, Erwan Bimbard, Rajiv Boddeda, Imam Usmani, Alexei
  Ourjoumtsev, and Philippe Grangier.
\newblock Quantum statistics of light transmitted through an intracavity
  rydberg medium.
\newblock {\em New Journal of Physics}, 16(4):043020, 2014.

\bibitem{jia2021span}
Feng-Dong Jia, Xiu-Bin Liu, Jiong Mei, Yong-Hong Yu, Huai-Yu Zhang, Zhao-Qing
  Lin, Hai-Yue Dong, Jian Zhang, Feng Xie, and Zhi-Ping Zhong.
\newblock Span shift and extension of quantum microwave electrometry with
  rydberg atoms dressed by an auxiliary microwave field.
\newblock {\em Physical Review A}, 103(6):063113, 2021.

\bibitem{main1994rydberg}
J{\"o}rg Main and G{\"u}nter Wunner.
\newblock Rydberg atoms in external fields as an example of open quantum
  systems with classical chaos.
\newblock {\em Journal of Physics B: Atomic, Molecular and Optical Physics},
  27(13):2835, 1994.

\bibitem{dunning2009engineering}
FB~Dunning, JJ~Mestayer, Carlos~O Reinhold, S~Yoshida, and J~Burgd{\"o}rfer.
\newblock Engineering atomic rydberg states with pulsed electric fields.
\newblock {\em Journal of Physics B: Atomic, Molecular and Optical Physics},
  42(2):022001, 2009.

\bibitem{holloway2018quantum}
Christopher~L Holloway, Matthew~T Simons, Marcus~D Kautz, Abdulaziz~H Haddab,
  Joshua~A Gordon, and Thomas~P Crowley.
\newblock A quantum-based power standard: Using rydberg atoms for a
  si-traceable radio-frequency power measurement technique in rectangular
  waveguides.
\newblock {\em Applied Physics Letters}, 113(9), 2018.

\bibitem{shirley1965solution}
Jon~H Shirley.
\newblock Solution of the schr{\"o}dinger equation with a hamiltonian periodic
  in time.
\newblock {\em Physical Review}, 138(4B):B979, 1965.

\bibitem{dion1976time}
David~R Dion and Joseph~O Hirschfelder.
\newblock Time-dependent perturbation of a two-state quantum system by a
  sinusoidal field.
\newblock {\em In: Advances in chemical physics. Volume 35. New York},
  35:265--350, 1976.

\bibitem{meyer2020assessment}
David~H Meyer, Zachary~A Castillo, Kevin~C Cox, and Paul~D Kunz.
\newblock Assessment of rydberg atoms for wideband electric field sensing.
\newblock {\em Journal of Physics B: Atomic, Molecular and Optical Physics},
  53(3):034001, 2020.

\bibitem{ho1983semiclassical}
Tak-San Ho, Shih-I Chu, and James~V Tietz.
\newblock Semiclassical many-mode floquet theory.
\newblock {\em Chemical Physics Letters}, 96(4):464--471, 1983.

\bibitem{ho1984semiclassical2}
Tak-San Ho and Shih-I Chu.
\newblock Semiclassical many-mode floquet theory. ii. non-linear multiphoton
  dynamics of a two-level system in a strong bichromatic field.
\newblock {\em Journal of Physics B: Atomic and Molecular Physics},
  17(10):2101, 1984.

\bibitem{ho1985semiclassical3}
Tak-San Ho and Shih-I Chu.
\newblock Semiclassical many-mode floquet theory. iii. su (3) dynamical
  evolution of three-level systems in intense bichromatic fields.
\newblock {\em Physical Review A}, 31(2):659, 1985.

\bibitem{ho1985semiclassical4}
Tak-San Ho and Shih-I Chu.
\newblock Semiclassical many-mode floquet theory. iv. coherent population
  trapping and su (3) dynamical evolution of dissipative three-level systems in
  intense bichromatic fields.
\newblock {\em Physical Review A}, 32(1):377, 1985.

\bibitem{saffman2010quantum}
Mark Saffman, Thad~G Walker, and Klaus M{\o}lmer.
\newblock Quantum information with rydberg atoms.
\newblock {\em Reviews of modern physics}, 82(3):2313, 2010.

\bibitem{weimer2010rydberg}
Hendrik Weimer, Markus M{\"u}ller, Igor Lesanovsky, Peter Zoller, and
  Hans~Peter B{\"u}chler.
\newblock A rydberg quantum simulator.
\newblock {\em Nature Physics}, 6(5):382--388, 2010.

\bibitem{lee2012collective}
Tony~E Lee, Hartmut Haeffner, and MC~Cross.
\newblock Collective quantum jumps of rydberg atoms.
\newblock {\em Physical review letters}, 108(2):023602, 2012.

\bibitem{saffman2016quantum}
Mark Saffman.
\newblock Quantum computing with atomic qubits and rydberg interactions:
  progress and challenges.
\newblock {\em Journal of Physics B: Atomic, Molecular and Optical Physics},
  49(20):202001, 2016.

\bibitem{adams2019rydberg}
Charles~S Adams, Jonathan~D Pritchard, and James~P Shaffer.
\newblock Rydberg atom quantum technologies.
\newblock {\em Journal of Physics B: Atomic, Molecular and Optical Physics},
  53(1):012002, 2019.

\bibitem{cox2018quantum}
Kevin~C Cox, David~H Meyer, Fredrik~K Fatemi, and Paul~D Kunz.
\newblock Quantum-limited atomic receiver in the electrically small regime.
\newblock {\em Physical Review Letters}, 121(11):110502, 2018.

\end{thebibliography}
\newpage

\end{document}